\documentclass[sigconf,9pt]{acmart}
\usepackage[normalem]{ulem}
\usepackage{graphicx}
\usepackage{amsmath}
\usepackage{verbatim}
\usepackage{multirow}
\usepackage{setspace}
\usepackage{soul}
\usepackage{color}
\usepackage{tikz}
\usepackage{url}
\usepackage{subfig}
\usepackage{float}
\usepackage[linesnumbered,ruled]{algorithm2e}

\def\name{ECStream}
\def\dataratio{73.01\%}
\def\latencyratio{20.37\%}

\renewcommand\footnotetextcopyrightpermission[1]{} 
\acmConference[]{arXiv preprint}{2023}{online}

\begin{document}


\title{Edge Cloud Collaborative Stream Computing for Real-Time Structural Health Monitoring}

\author{Wenzhao Zhang, Cheng Guo, Yi Gao, and Wei Dong}
\affiliation{%
	\institution{College of Computer Science, Zhejiang University}
	\streetaddress{38 Zheda Rd}
	\city{Hangzhou}
	\country{China}}
\email{{wz.zhang, guo.cheng, gaoy, dongw}@zju.edu.cn}

\begin{abstract}
Structural Health Monitoring (SHM) is crucial for the safety and maintenance of various infrastructures. Due to the large amount of data generated by numerous sensors and the high real-time requirements of many applications, SHM poses significant challenges. Although the cloud-centric stream computing paradigm opens new opportunities for real-time data processing, it consumes too much network bandwidth. In this paper, we propose ECStream, an Edge Cloud collaborative fine-grained stream operator scheduling framework for SHM. We collectively consider atomic and composite operators together with their iterative computability to model and formalize the problem of minimizing bandwidth usage and end-to-end operator processing latency. Preliminary evaluation results show that ECStream can effectively balance bandwidth usage and end-to-end operator computation latency, reducing bandwidth usage by 73.01\% and latency by 34.08\% on average compared to the cloud-centric approach.
\end{abstract}



\maketitle

\section{Introduction} \label{sec:intro}
Structural Health Monitoring (SHM) is crucial for the safety and maintenance of various infrastructures such as bridges, buildings, and dams \cite{chang2003health, farrar2007introduction, brownjohn2007structural}. 
Due to the large amount of data generated by numerous sensors and high real-time requirements of many applications, SHM poses significant challenges. 
For instance, detecting cracks in bridges or monitoring vibrations in wind turbines require timely and accurate analysis of massive data streams.

Traditional bridge SHM typically relies on cloud-centric batch data processing \cite{zonzini2020structural, mishra2022structural, bhuiyan2016secured}, where all raw data is transmitted to the cloud and processed after accumulating a certain amount. 
However, this approach results in poor timeliness for various applications. 

Recently, distributed big data processing frameworks, such as Spark and Flink, have been developed to enable real-time stream processing. 
These frameworks have been widely applied in fields like finance, creating opportunities for improving SHM.

However, large-scale infrastructure SHM systems often employ numerous IoT sensors, generating massive amount of data \cite{lynch2007overview}. 
The existing stream processing frameworks are usually deployed on the cloud to obtain stable performance \cite{ishii2011elastic, ta2016big, neumeyer2010s4}.
The limited bandwidth between cloud and IoT devices, along with network link transmission uncertainty, can lead to significant end-to-end latency. 
Fortunately, the increasing adoption of edge devices (e.g., industrial control systems) create opportunities for low-latency data-processing.
Recent works have exploit the benefit of edge-centric stream computing paradigm \cite{fu2019edgewise, liu2021dart, xu2022amnis} or build specialized toolkit for it \cite{battulga2021speck}.

Nevertheless, edge devices usually equip with limited computing resources while stream computing requires large volume of memory to maintain stable performance \cite{han2015spark, zou2022assasin}.
As a result, the edge-centric stream computing paradigm can lead to sub-optimal performance especially in terms of end-to-end latency.

One straightforward way to circumvent this problem is to use cloud-edge collaborative stream processing, which unfortunately, faces non-trivial challenges especially under SHM scenarios. 

Specifically, SHM involves numerous applications composed of processing units, such as data statistical metrics calculations (mean, variance, etc.), Fourier transforms, Hilbert transforms, wavelet transforms, and power spectral density calculations.
Although we can directly encapsulate these units into stream operators, it is hard to capture the fine-grained dependencies and iterative computability among them, leading to sub-optimal solutions.

\begin{figure}
	\centering
	\includegraphics[scale=0.52]{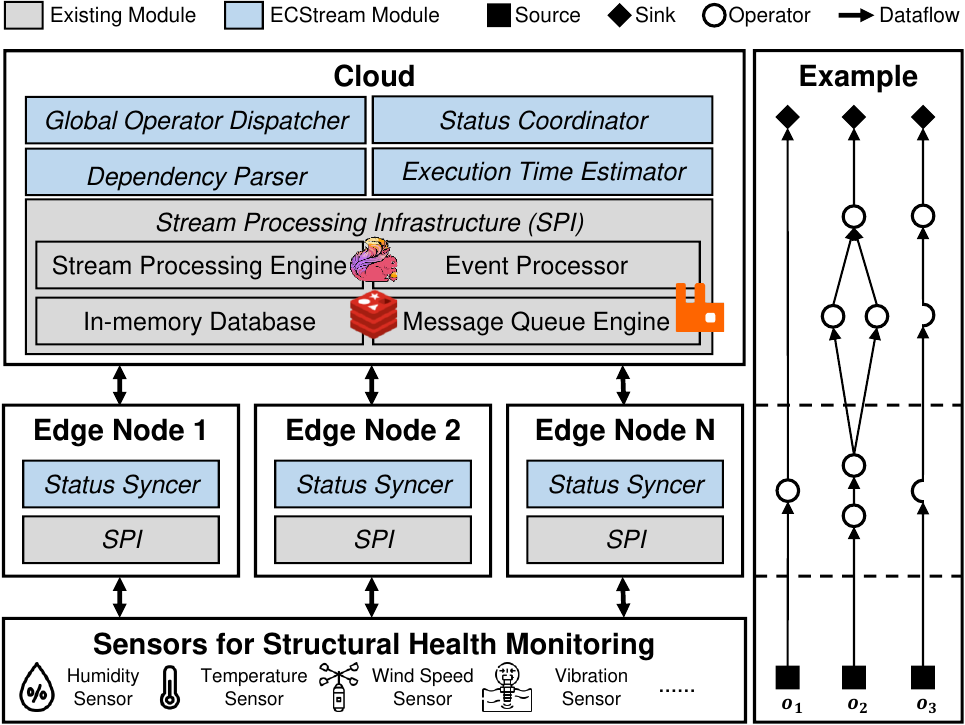}
	\caption{\name\ overview and an example.}
	\label{fig:overview}
\end{figure}

The right part of Figure \ref{fig:overview} shows an illustrating example.
$o_1$ represents the most ordinary kind of operators that do not depend on other operators. 
We call this kind atomic operators.
$o_2$ represents the other kind of operators that depend on other operators.
We call this kind composite operators.
For example, calculating wind deviation angles requires the mean values of the x and y directions, followed by further trigonometric calculations. 
Besides atomic and composite operators, there is another classification methodology, i.e., whether or not the operator is iteratively computable, as indicated by $o_3$.
One typical example of this kind is the ``mean'' function.
Concretely, we can split the calculation of the mean of 100 data points up in a 50-50 manner (or other arbitrary ratio) with no loss of accuracy.


To address the above challenge, we propose \name, an \underline{E}dge \underline{C}loud collaborative fine-grained \underline{S}tream operator scheduling framework for SHM (the architecture is shown as the left part of Figure \ref{fig:overview}). 
We collectively consider atomic and composite operators together with their iterative computability to model and formalize the problem of minimizing bandwidth usage and end-to-end operator processing latency.
We implemented \name\ and evaluated its performance using large-scale real-world data traces from Hong Kong–Zhuhai–Macao Bridge and simulations. 
Results show that \name\ can effectively balance bandwidth usage and end-to-end operator computation latency, reducing bandwidth usage by \dataratio\ and latency by \latencyratio\ on average.

\section{Problem Statement and Solution} \label{sec:statement}
\subsection{Problem Statement}
\quad\ We first introduce the following notations.

\begin{itemize}
	\item $\mathbb{O}, \mathbb{S}, \mathbb{N}$. The set of stream operators, logical sensors, and edge computing nodes.
	\item $i,j,k,o_i,s_j,n_k,l_{ijk}$. 
	We use $o_i$, $s_j$, and $n_k$ to denote a specific stream operator, a sensor, and an edge node, respectively. 
	Here, $o_i$ is uniquely defined by a seven-entry tuple -- $<\mathbb{S}_i, \mathbb{O}_i, func_i, iter_i, window_i, step_i, freq_i>$, where $\mathbb{S}_i$ is the logical sensor(s) that $o_i$ relies on;
	$\mathbb{O}_i$ is the operator(s) that $o_i$ consumes, indicating that $o_i$ is a composite operator if $|\mathbb{O}_i| \geq 1$.
	$func_i$ is the transformation/aggregation function; $window_i$, $step_i$, and $freq_i$ indicate the window size, the sliding step, and the output frequency of $o_i$, respectively.
	$iter_i$ is an indicator to specifies if $o_i$ is iteratively computable ($iter_i = 1$) or not ($iter_i = 0$).
	A sensor $s_j$, on the other hand, is uniquely wired to an edge node $n_k$.
	Typically, we use $l_{ijk}$ as an indicator variable to denote whether $s_j$ of $o_i$ is at $n_k$ ($l_{ijk} = 1$) or not ($l_{ijk} = 0$).
	\item $C, M, B, D, T$. 
	We use them to denote the volume of CPU cycles, memory, bandwidth, transmitted data, and latency, respectively.
	\item $\gamma$. We use $\gamma_{ij}$ to denote that we will offload a specific ratio of data and computation of $s_{ij}$ to the cloud.
\end{itemize}

With the above definitions, the stream operator dispatching and scheduling problem can be formulated as follows with the optimization criterion being the trade-off value of total bandwidth consumption and end-to-end latency.

\centerline{Find the values of $\gamma_{ij}$ ($\forall i \in |\mathbb{O}|, \forall j \in |\mathbb{S}_i|$)}
\vspace{-3.5ex}
{
	\small
	\begin{align}
		\min &\sum_{\forall i \in |\mathbb{O}|}\sum_{\forall k \in |\mathbb{N}|} D_{ik}(\gamma_{ij})  \notag\\
		\textbf{s.t.}~~&\gamma_{ij} \in [0,1], \forall i \in |\mathbb{O}|, \forall j \in |\mathbb{S}_i|, if~iter_i=1 \label{eq:iter}\\
		& \gamma_{ij} \in \{0, 1\}, \forall i \in |\mathbb{O}|, \forall j \in |\mathbb{S}_i|, if~iter_i=0 \label{eq:niter}\\
		& \gamma_{ij_1} = \gamma_{ij_2}, \forall i \in |\mathbb{O}|, \forall j_1, j_2 \in |\mathbb{S}_i| \label{eq:sensor}\\
		& l_{ijk} \in \{0, 1\}, \forall i \in |\mathbb{O}|, \forall j \in |\mathbb{S}_i|, \forall k \in |\mathbb{N}_{ij}| \label{eq:locvar}\\
		& \sum_{\forall k \in |\mathbb{N}_{ij}|} l_{ijk} = 1, \forall i \in |\mathbb{O}|, \forall j \in |\mathbb{S}_i| \label{eq:locsum}\\
		& \gamma_{ij} \geq \lfloor \frac{1}{|\mathbb{S}_{i}|}\sum_{\forall j \in |\mathbb{S}_{i}|}\prod_{\forall k \in |\mathbb{N}_{ij}|} l_{ijk} \rfloor \label{eq:diffloc}\\
		& \gamma_{ij} \geq \lfloor \frac{1}{|\mathbb{O}_{i}|} \sum_{\forall i' \in |\mathbb{O}_i|} \frac{1}{|\mathbb{S}_{i'}|}\sum_{\forall j \in |\mathbb{S}_{i'}|}\prod_{\forall k \in |\mathbb{N}_{i'j}|} l_{i'jk} \rfloor, if~|\mathbb{O}_i| > 0 \label{eq:comp_diffloc}\\
		& \gamma_{ij} \geq 1 - \lfloor \frac{1}{|\mathbb{S}_{i'}|} \sum_{\forall i' \in |\mathbb{O}_i|} \gamma_{i'}\rfloor, if~|\mathbb{O}_i| > 0 \label{eq:comp_iter}\\
		& \gamma_{ij} = \min_{\forall i' \in |\mathbb{O}_i|} \gamma_{i'j}, if~|\mathbb{O}_i| > 0 \label{eq:min}\\
		& T_i(\gamma_{ij}) \leq T_i^{req}, \forall i \in |\mathbb{O}| \label{eq:t} \\
		& \sum_{\forall i \in |\mathbb{O}|} C_{ik}(\gamma_{ij}) < C_k, \forall k \in |\mathbb{N}| \label{eq:cpu} \\
		& \sum_{\forall i \in |\mathbb{O}|} M_{ik}(\gamma_{ij}) < M_k, \forall k \in |\mathbb{N}| \label{eq:mem}
	\end{align}
}

In the following, we further give the detailed definition of $D_{ik}(\gamma_{ij})$, $\gamma_j$, $T_i(\gamma_{ij})$, $C_{ik}(\gamma_{ij})$, and $M_{ik}(\gamma_{ij})$.

$D_{ik}(\gamma_{ij})$ can be divided into three parts, the raw sensor data, the intermediate, and final results, as shown in Eq. (\ref{eq:d}).
{
	\small
	\begin{align}
		D_{ik}(\gamma_{ij}) = \sum_{\forall j \in |\mathbb{S}_i|} D_{ijk} \cdot \gamma_{j} + (\lceil \gamma_{ij} \rceil - \lfloor \gamma_{ij} \rfloor) \cdot D_i^{int} + \lfloor 1-\gamma_{ij} \rfloor \cdot D_i^{res} \label{eq:d}
	\end{align}
}

Unlike $\gamma_{ij}$ which is defined over an operator, the system actually needs the $\gamma_j$ that is defined over a (logical) sensor to truly decide which portion of the data should be processed locally at the edge while others need to be offloaded.
The derivation from $\gamma_{i}$ to $\gamma_{j}$ is relatively straightforward by obtaining the maximum value of each $\gamma_{ij}$, which can be formulated as Eq. (\ref{eq:r}).
{
	\small
	\begin{align}
		\gamma_{j} = \max_{\forall i \in |\mathbb{O}|} \gamma_{ij}, \forall j \in |\mathbb{S}| \label{eq:r}
	\end{align}
}

$T_i(\gamma_{ij})$ can be divided into four parts, the processed time at the edge, the data transmission time between edge and cloud, the waiting time, and the processed time at the cloud, which can be formally defined in Eq. (\ref{eq:dft}).
{
	\small
	\begin{align}
		T_i(\gamma_{ij}) = T_i^{edge}(\gamma_{ij}) + T_i^{trans}(\gamma_{ij}) + T_i^{wait}(\gamma_{ij}) + T_i^{cloud}(\gamma_{ij}) \label{eq:dft}
	\end{align}
}

Each $T_i^{edge}(\gamma_{ij})$ can be derived from the maximum value of all $k$ edge nodes as shown in Eq. (\ref{eq:edgemax}).
{
	\small
	\begin{align}
		T_i^{edge}(\gamma_{ij}) = \max_{k \in |\mathbb{N}|} T_{ik}^{edge}(\gamma_{ij}) \label{eq:edgemax}
	\end{align}
}

Each $T_{ik}^{edge}(\gamma_{ij})$ is calculated as follows.
{
	\small
	\begin{align}
		T_{ik}^{edge}(\gamma_{ij}) = \frac{\sum_{\forall j \in |\mathbb{S}_i|} C_{ijk} \cdot \gamma_{ij}}{C_k^{unit}}
	\end{align}
}

We can also define $T_i^{trans}(\gamma_{ij})$ similar to $T_i^{edge}(\gamma_{ij})$.
{
	\small
	\begin{align}
		T_i^{trans}(\gamma_{ij}) = \max_{k \in |\mathbb{N}|} T_{ik}^{trans}(\gamma_{ij})
	\end{align}
}

Each $T_{ik}^{trans}(\gamma_{ij})$, on the other hand, is determined by the transmitted data volume and the corresponding bandwidth.
{
	\small
	\begin{align}
		T_{ik}^{trans}(\gamma_{ij}) = \frac{D_{ik}(\gamma_{ij})}{B_k}
	\end{align}
}

$T_i^{wait}(\gamma_{ij})$ is a little more complicated.
If $o_i$ is not composite, there is no need to wait results from others so it is set to 0.
Otherwise, it will be set to the maximum deviation of its dependent operators.
Eq. (\ref{eq:twait}) formally specifies the aforementioned definition.
{
	\small
	\begin{align}
		T_i^{wait}(\gamma_{ij}) = 
		\begin{cases}
			0, if~|\mathbb{O}_i| = 0\\
			\max_{\forall i'_1, i'_2 \in |\mathbb{O}_i|} T_{i'_1}(\gamma_{i'_1j}) - T_{i'_2}(\gamma_{i'_2j}), if~|\mathbb{O}_i| > 0
		\end{cases}
		\label{eq:twait}
	\end{align}
}

Similar to $T_{ik}^{edge}(\gamma_{ij})$, we can define $T_i^{cloud}(\gamma_{ij})$ as follows.
{
	\small
	\begin{align}
		T_i^{cloud}(\gamma_{ij}) = \frac{[\sum_{\forall j \in |\mathbb{S}_i|} C_{ij} \cdot (1-\gamma_{ij})] + C_{i}^{res}}{C_{cloud}^{unit}}
	\end{align}
}

We can obtain $C_{ik}(\gamma_{ij})$ and $M_{ik}(\gamma_{ij})$ by summing up all resources consumed by co-located operations.
{
	\small
	\begin{align}
		C_{ik}(\gamma_{ij}) = \sum_{\forall j \in |\mathbb{S}_i|} C_{ijk} \cdot \gamma_{ij}
		\label{eq:c}
	\end{align}
}

{
	\small
	\begin{align}
		M_{ik}(\gamma_{ij}) = \sum_{\forall j \in |\mathbb{S}_i|} M_{ijk} \cdot \gamma_{ij}
		\label{eq:m}
	\end{align}
}

Eq. (\ref{eq:iter})\textasciitilde(\ref{eq:mem}) are constraints and we use C-$x$ to denote them.
C-\ref{eq:iter} means that if $iter_i=1$, $\gamma_{ij}$ can be set to a continuous value within [0,1].
C-\ref{eq:niter} holds when $iter_i=0$, in this case, $\gamma_{ij}$ can be set either to 0 or 1. 
C-\ref{eq:sensor} is valid when $o_i$ requires multiple (logical) sensors, under this circumstance, we impose that the $\gamma$ for all $j$ sensor of $o_i$ should be the same as otherwise $o_i$ cannot be calculated directly because of data shortage.
C-\ref{eq:locvar} constraints the value of $l_{ijk}$ to \{0,1\} while C-\ref{eq:locsum} specifies that $s_j$ of $o_i$ can only locates on a specific $n_k$. 
C-\ref{eq:diffloc} takes effect when $o_i$ requires multiple (logical) sensors, if these sensors spread over different nodes, $\gamma_{ij}$ has to set to 1, i.e., $o_i$ should be calculated on the cloud because only there can $o_i$ obtain all the data.
C-\ref{eq:comp_diffloc} is similar to C-\ref{eq:diffloc} except that $o_i$ in this case is composite.
C-\ref{eq:comp_iter} also applies for the cases when $o_i$ is composite, which means that if one of the dependent $o_{i'}$ is iteratively computable and has already decided to be partially computed on edge/cloud side, $o_i$ has to be processed on the cloud.
Here we assume that the (intermediate) results on the cloud will not be transferred back to the edge.
C-\ref{eq:min} specifies that the $\gamma_{ij}$ of a composite $o_i$ should be set to the minimal value of its dependent $o_{i'}$ because that decides the maximum volume of ready-to-be-processed data.
C-\ref{eq:t}, C-\ref{eq:cpu}, and C-\ref{eq:mem} denote that the end-to-end latency, CPU, and memory should meet the user specify or physical capacities.

\subsection{Solution} 
To solve the problem, one straightforward way is to exhaustively enumerate all combinations for $\gamma_{ij}$, which is computation intensive.
We observe that the determination of $\gamma_{ij}$ for composite operators highly relies on that of their dependent atomic operators.
This opens the opportunity for reducing the enumerations by progressively determining $\gamma_{ij}$ first the atomic then the composite operators.
We argue that this procedure will make no difference to the final optimal results because of the constraint effect of C-\ref{eq:sensor}, C-\ref{eq:min}, and Eq. (\ref{eq:r}). 
Furthermore, we also apply pre-flight checks for C-\ref{eq:t}, C-\ref{eq:cpu}, and C-\ref{eq:mem} to enable early-exit, which will further reduce the enumeration overhead.
Algorithm \ref{algo:two-stage} summaries the overall workflow.

\begin{algorithm}[t]
	\caption{Two-stage progressive solving algorithm}
	\label{algo:two-stage}
	\SetKwData{In}{\textbf{in}}
	\SetKwData{To}{\textbf{to}}
	\SetKwProg{Fn}{Cascade}{:}{}
	\DontPrintSemicolon
	\SetAlgoVlined
	\KwIn {$\mathbb{O}, T^{req}, profile(\mathbb{O}), D, C, M, B, \Delta$}
	\KwOut {$\gamma_{ij}$}
	\Begin{
		$\mathbb{O}\gets reorder(\mathbb{O})$
		
		\Fn{{$loops~of~all~o_i \in \mathbb{O}$}}{
			$initialize~optimalD$
			
			\For{$\gamma_{i} \gets 0$ \KwTo $1$}{
				$initialize~D_i, C_i^{edge}, C_i^{cloud}, M_i^{edge}, M_i^{cloud}, T_i$
				
				\For{$j \in \mathbb{S}_i$ }{
					$update~D_i, C_i^{edge}, C_i^{cloud}, M_i^{edge}, M_i^{cloud}$
					
					\If{\text{\textbf{not}}~$pre\_flight\_check\_resource()$}{
						\text{\textbf{break}}\;
					}
					
					\If{\text{\textbf{not}}~$pre\_flight\_check\_optimalD()$}{
						\text{\textbf{break}}\;
					}
					
					$update~T_i~as~Eq.~(\ref{eq:dft})$
					
					\If{\text{\textbf{not}}~$pre\_flight\_check\_latency()$}{
						\text{\textbf{break}}\;
					}
					
					$record~\gamma_{i}$
				}
				$\gamma_{i} \gets \gamma_{i} + \Delta$\;
			}
			
			\tcc{the following code will only be executed at the most inner loop}
			
			\If{\text{\textbf{not}}~$final\_check()$} 
			{
				\text{\textbf{break}}\;
			}
			
			\If{$sum(\forall D_i)<optimalD$}
			{
				$update~optimalD$
			}
		}
		
		\Return{$\gamma_{ij}$}
	}
\end{algorithm}

The input of the algorithm includes four aspects.
First, the operator set $\mathbb{O}$, their end-to-end latency requirement $T^{req}$, and their profile, which mainly contains $C_{ik}$, $M_{ik}$, and $D_i^{int}$ for each $o_i$.
Second, the data volume $D$ of each sensor.
Third, the specifications of the cloud and edge nodes, including the capacity and unit of CPU in $C$, memory capacity $M$, as well as the bandwidth between edge node and the cloud $B$.
Fourth, the $\gamma_{ij}$ searching step $\Delta$.
Given the input, the algorithm outputs $\gamma_{ij}$.

The algorithm basically has two parts.
The first part reorder the $\mathbb{O}$ (Line 2), making sure that the algorithm will first go through the atomic operators then the composite ones.
The second part enumerate possible combinations with three rounds of pre-flight checks (Lines 9-10, 11-12, 14-15), which will trigger early-exit if the corresponding checks fail.
During the second part, the algorithm will update intermediate results of $D_i$, $C_i$, $M_i$, and $T_i$ according to Eq. (\ref{eq:d}), Eq. (\ref{eq:c}), Eq. (\ref{eq:m}), and Eq. (\ref{eq:dft}).
The algorithm will update $optimalD$ at the most inner loop and finally return $\gamma_{ij}$ after all enumerations are finished.

\begin{table*}[]
	\centering
	\caption{Specifications of the evaluated operators.
	cc: cross-correlation, avgws: average wind speed, avgwa: average wind angle, gf: gust factor, fws: fluctuating wind speed, ti: turbulence intensity, aoa: angle of attack, awd: angle of wind deflection.
	Each item within the brace (i.e., ``\{...\}'') indicates the $(window, step, freq)$ has the same value, e.g., \{600\} represents (600, 600, 600).
	}
	\label{tab:op}
	\resizebox{\linewidth}{!}{%
		\begin{tabular}{|c|c|c|c|c|c|}
			\hline
			\textbf{No.}   & \textbf{$\mathbb{S}$}                                           & \textbf{$\mathbb{O}$}    & \textbf{$func$}   & \textbf{$iter$} & \textbf{$(window, step, freq)$}       \\ \hline
			1-4   & RHS, TMP, UAN, ULT, WIM, VIB, VIC, RSG, HPT, DPM, GPS & Null         & mean   & $\surd$    & \{60, 600, 3600, 86400\} \\ \hline
			5-8   & RHS, TMP, UAN, ULT, WIM, VIB, VIC, RSG, HPT, DPM, GPS & Null         & msqrt  & $\surd$    & \{60, 600, 3600, 86400\} \\ \hline
			9-12  & RHS, TMP, UAN, ULT, WIM, VIB, VIC, RSG, HPT, DPM, GPS & Null         & max    & $\times$    & \{60, 600, 3600, 86400\} \\ \hline
			13-16 & RHS, TMP, UAN, ULT, WIM, VIB, VIC, RSG, HPT, DPM, GPS & Null         & min    & $\times$    & \{60, 600, 3600, 86400\} \\ \hline
			17-20 & RHS, TMP, UAN, ULT, WIM, VIB, VIC, RSG, HPT, DPM, GPS & Null         & first  & $\times$    & \{60, 600, 3600, 86400\} \\ \hline
			21-24 & RHS, TMP, UAN, ULT, WIM, VIB, VIC, RSG, HPT, DPM, GPS & Null         & last   & $\times$    & \{60, 600, 3600, 86400\} \\ \hline
			25-28 & RHS, TMP, UAN, ULT, WIM, VIB, VIC, RSG, HPT, DPM, GPS & Null         & range  & $\times$    & \{60, 600, 3600, 86400\} \\ \hline
			29-32 & RHS, TMP, UAN, ULT, WIM, VIB, VIC, RSG, HPT, DPM, GPS & Null         & std    & $\surd$    & \{60, 600, 3600, 86400\} \\ \hline
			33-36 & RHS, TMP, UAN, ULT, WIM, VIB, VIC, RSG, HPT, DPM, GPS & Null         & var    & $\surd$    & \{60, 600, 3600, 86400\} \\ \hline
			37-40 & HPT, VIB, VIC, RSG, RHS, UAN, TMP, DPM, GPS, ULT, FLX & Null         & cov    & $\surd$    & \{60, 600, 3600, 86400\} \\ \hline
			41-44 & GPS, HPT, DPM                                 & {[}49-52{]}  & speed  & $\surd$    & \{60, 600, 3600, 86400\} \\ \hline
			45-48 & GPS, HPT, DPM                                 & {[}41-44{]}  & acc    & $\surd$    & \{60, 600, 3600, 86400\} \\ \hline
			49-52 & GPS, HPT, DPM                                 & Null         & disp   & $\surd$    & \{60, 600, 3600, 86400\} \\ \hline
			53    & HPT, VIB, VIC, RSG, RHS, UAN, TMP, DPM, GPS, ULT, FLX & Null         & cc     & $\times$    & (900, 300, 100)          \\ \hline
			54    & HPT, VIB, VIC, RSG, RHS, UAN, TMP, DPM, GPS, ULT, FLX & Null         & filter & $\times$    & (10, 10, 10)             \\ \hline
			55    & HPT, VIB, VIC, RSG, RHS, UAN, TMP, DPM, GPS, ULT, FLX & {[}2{]}      & trend  & $\surd$    & (600, 60, 60)            \\ \hline
			56    & HPT, VIB, VIC, RSG, RHS, UAN, TMP, DPM, GPS, ULT, FLX & Null         & surge  & $\surd$    & (10, 1, 1)               \\ \hline
			57    & UAN                                         & {[}2{]}      & avgws  & $\surd$    & \{600\}                  \\ \hline
			58    & UAN                                         & {[}2, 57{]}  & avgwa  & $\times$    & \{600\}                  \\ \hline
			59    & UAN                                         & {[}2{]}      & gf     & $\surd$    & (3, 1, 1)                \\ \hline
			60    & UAN                                         & {[}57, 58{]} & fws    & $\times$    & \{1\}                    \\ \hline
			61    & UAN                                         & {[}57, 60{]} & ti     & $\times$    & \{600\}                  \\ \hline
			62    & UAN                                         & {[}2, 57{]}  & aoa    & $\surd$    & \{600\}                  \\ \hline
			63    & UAN                                         & {[}2{]}      & awd    & $\surd$    & \{600\}                  \\ \hline
	\end{tabular}
	}
\end{table*}

\begin{figure*}[ht]
	\centering
	\includegraphics[width=\linewidth]{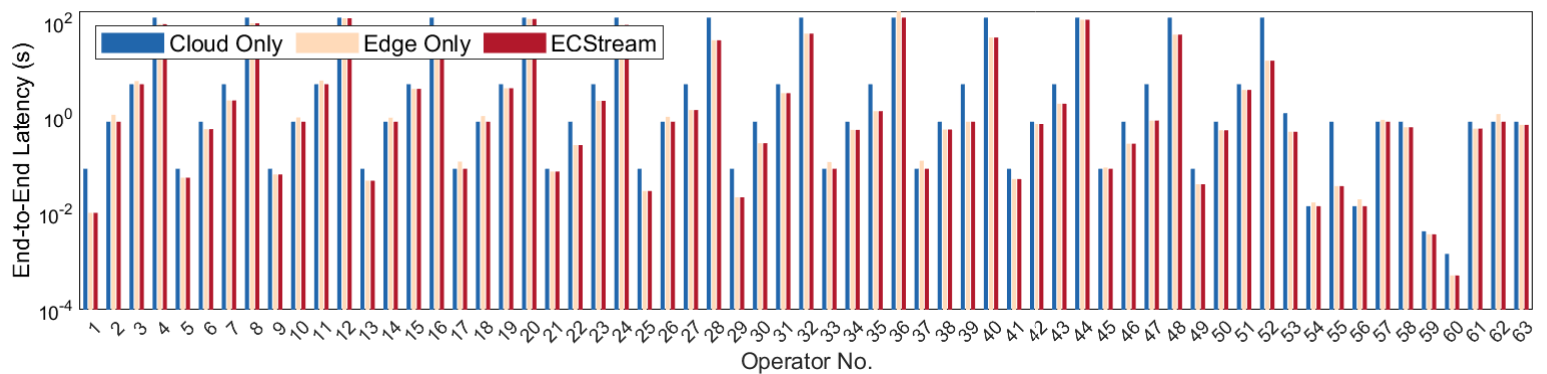}
	\caption{\name\ performance w.r.t end-to-end latency of different operators.}
	\label{fig:performance}
\end{figure*}

\section{Preliminary Evaluation}
In our preliminary evaluation, we analyze the performance and efficiency of \name\ in a trace-driven manner.
\subsection{Evaluation Setup}
\quad\textbf{Data trace.}
We use a large-scale real-world data trace come from Hong Kong–Zhuhai–Macao Bridge, which contains 1000+ data channels of 12 distinct types of sensors.
Each sensor has a sampling rate of 10 Hz.
We use the stream operator from the current structural health monitoring system of Hong Kong–Zhuhai–Macao Bridge with a total number of 63.
Table \ref{tab:op} shows the specifications of these operators, including their dependent sensor set ($\mathbb{S}$), dependent operator set ($\mathbb{O}$), calculation function ($func$), interative computability ($iter$), window ($window$), step ($step$), and query frequency ($freq$).
Here, $\mathbb{O}_i=Null$ is equivalent to $|\mathbb{O}_i|=0$, meaning that $o_i$ is atomic, otherwise, it is composite. 

\textbf{Baseline.}
We mainly compare two baselines as follows:
\begin{itemize}
	\item Cloud-Only (CO) \cite{ishii2011elastic, ta2016big, neumeyer2010s4}: 
	Upload all the raw sensor data directly to the cloud without caching and run the stream processing on the cloud.
	\item Edge-Only (EO) \cite{fu2019edgewise, liu2021dart, xu2022amnis}:
	Keep all the raw sensor data, run the stream processing at the edge, and only upload the results to the cloud.
\end{itemize}

\textbf{Platform specifications and operator profiling.}
We run our on two devices.
A PC that equips with an Intel Core i7-8700 CPU @3.2 GHz and 32 GB DDR4 memory @2666 MHz, which acts as the cloud.
A Raspberry Pi 4B that equips with ARM Cortex-A72 CPU @1.5 GHz and 8GB LPDDR4 memory @3200 MHz, which acts as the edge device.
The bandwidth between the edge and the cloud is set to the average of real-world speed of 4G network \cite{4g}, which is about 1.56 MB/s.
We run the operators on these two platforms and profile their resource consumption (e.g., CPU, memory, and bandwidth) and execution latency.

\textbf{Methodology.}
To evaluate the performance, given the operator specifications (as shown in Table \ref{tab:op}) and profile, we simulate the continuous data stream and record the according data transmission volume as well as end-to-end latency of both baselines and \name.
Each logical sensor is assumed to be connected to one specific edge node.
The time constraint of each operator $T^{req}_i$ is set to 10\% larger than the average execution time of $T^{cloud}_i$.

\begin{figure}[]
	\centering
	\includegraphics[width=0.9\linewidth]{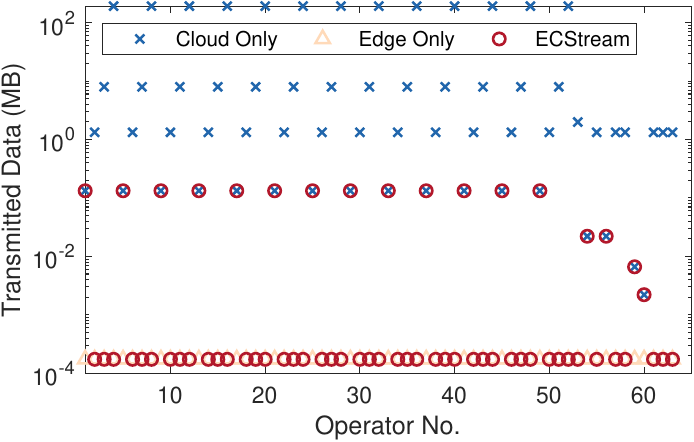}
	\caption{\name\ performance w.r.t transmitted data of different operators.}
	\label{fig:data}
\end{figure}

\subsection{Main Results}
\quad\textbf{Performance.}
Figure \ref{fig:performance} shows the end-to-end latency comparison.
We can see that \name\ has a better performance than CO and EO for all the 63 operators and a low data transmission volume.
Overall, \name\ can reduce an average of \latencyratio\ end-to-end latency compared to both CO and EO.
Figure \ref{fig:data} shows the transmitted data comparison.
\name\ can reduce \dataratio\ data volume compared to CO.
The main reason \name\ can outperform baselines is that it leverage the computation resources of both cloud and edge side and carefully trade-off data transmission volume and end-to-end latency with the help of our fine-grained stream operator modeling.

\section{Conclusion and Future Work}
In this work, we present \name, the first cloud-edge cooperative stream processing framework for SHM.
We build a prototype of \name\ and conduct evaluation on real-world data traces from Hong Kong–Zhuhai–Macao Bridge.
Compared to cloud-only and edge-only solutions, \name\ can reduce an average of \dataratio\ transmitted data and \latencyratio\ end-to-end latency.

We consider our future work in three aspects.
Firstly, \name\ currently treats operators with functions like Fourier and Hilbert transforms as not iteratively computable ones.
There are, however, some ways to translate them into iteratively computable with approximation methods. 
We would like to conduct a thorough survey over SHM operators and build an operator category together with a systematic approximation methodology to trade-off computation accuracy and efficiency.

Secondly, \name\ currently assumes that the network condition is stable.
In reality, however, the condition can be quite fluctuated because of the complex and highly changing environment of field deployment.
We would like to explore the performance of \name\ under uncertain 5G network and build a more robust system model in this situation.

Lastly, we evaluate \name\ with trace-driven simulation under hard assumptions (e.g., one edge node is only connected with one logical sensor).
We would like to: 
(1) refine the trace with real-world deployment topology and 
(2) build a testbed with production-ready open-source software and conduct real-world evaluation on it to further validate our design.

\section*{Acknowledgments}
This work is supported by the National Key R\&D Program of China under Grant No. 2019YFB1600700.

\bibliographystyle{plain}
\bibliography{references}

\end{document}